\documentclass[12pt]{iopart}

\newcommand\ep{\epsilon}

\newcommand\m{\mu}

\newcommand{\be}{\begin{equation}}
\newcommand{\ee}{\end{equation}}
\newcommand{\bea}{\begin{eqnarray}}
\newcommand{\eea}{\end{eqnarray}}
\newcommand{\ba}[1]{\begin{array}{#1}}
\newcommand{\ea}{\end{array}}

\newcommand{\diracslash}[1]{#1\llap{/\kern2pt}}

\begin{document}

\title{Recent developments in weak-coupling color superconductivity}

\author{Qun Wang
\footnote[3]{email:qwang@th.physik.uni-frankfurt.de} }

\address 
{Institut f\"ur Theoretische Physik,
J. W. Goethe-Universit\"at, D-60054 
Frankfurt am Main, Germany}

\begin{abstract}

Recent developments in weak-coupling color superconductivity 
are reviewed. These developments are as follows. The mean field gap equation 
is solved for most common superconducting phases up to subleading order; 
BCS relation is found to be violated in color-flavor-locking and 
color-spin-locking phases due to the two-gaps structure 
of the order parameter; The Debye and Meissner masses of 
gluons and photon with their rotated partners are calculated for these phases; 
We found that there is no electromagnetic Meissner effect 
in spin-one color superconductor; A proof of gauge parameter  
independence at subleading order is given in covariant gauge. 

\end{abstract}




It is widely accepted that the quark matter at low temperature 
and high density is a color superconductor. 
Quantum chromodynamics (QCD) tells us 
that the single-gluon exchange, which is attractive in 
the color-antitriplet channel \cite{bai84}, 
becomes the dominant interaction between quarks due to 
the aymptotic freedom \cite{asymp} in quark matter 
at very high density or large quark chemical potential $\m$. 
In recent years a lot of progress has been made 
in color superconductivity (for reviews, see e.g. \cite{review}). 
One of the most natural ways  
of studying color superconductivity, the so-called weak-coupling 
approach, is the QCD perturbation 
theory based on the assumption that the strong coupling constant is 
asymptotically small. The weak-coupling approach 
is valid only at very high density, but it is from the first principle and  
many insights can be drawn from it 
as guidelines to build up more phenomenological models. 
In this sense the weak-coupling approach is a unique and appealing 
perspective in viewing this marvelous phenomenon. 
There are a lot of fundamental issues in this framework. 
Here we review some of them which were already 
addressed and solved in the past few years in Frankfurt.

In a color superconductor with $N_f = 2$ massless flavors of quarks
(2SC phase), the value of the zero temperature gap at the 
Fermi surface is
\be \label{phi02SC}
\phi_0^{\rm 2SC}=2 \, \tilde{b} \, b_0'\, \mu\, \exp\left(
-\frac{\pi}{2\, \bar{g}}\right) \,\, ,
\ee
where
\be \label{constants}
\bar{g} \equiv \frac{g}{3 \sqrt{2}\, \pi} \,\, , \qquad
\tilde{b} \equiv 256 \pi^4 \left(\frac{2}{N_f g^2} \right)^{5/2}\,\, , \qquad
b_0' \equiv \exp\left(-\frac{\pi^2+4}{8}\right)
\,\, .
\ee
The term in the exponent of Eq.\ (\ref{phi02SC}) was
first computed in Ref.\ \cite{son99} and then confirmed in 
Ref.\ \cite{SchaferWilczek,pis00,hong,miransky}. 
It arises from the exchange of almost static magnetic gluons, 
which we call the leading contribution. 
The factor $\tilde{b}$ in front of the exponential originates from
the exchange of static electric and non-static magnetic gluons \cite{pis00,hsu00}. 
The prefactor $b_0'$ is due to the quark self-energy \cite{ren,wang}.
These prefactors are from subleading contributions in the 
gap equation.

In color superconductors, the mass shell of a quasiparticle
is determined by its excitation energy
\be \label{excite}
\ep _{k,r}(\phi)=\left[(k-\mu)^2+\lambda_r \,|\phi(\ep _{k,r},k)|^2
\right]^{1/2} \,\, ,
\ee
where $k\equiv |{\bf k}|$ is the modulus of the 3-momentum of the
quasiparticle, and $\phi(\ep _{k,r},k)$ is
the gap function on the quasiparticle mass shell. 
The index $r$ labels possible excitation branches in the superconductor,
which differ by the value of the constant $\lambda_r$.
At the Fermi surface, $k=\mu$, the true energy gap 
is $\sqrt{\lambda_r}\,\phi_0$. 
For example, in 2SC phase, quarks of two colors form Cooper 
pairs with total spin zero, while the third color
remains unpaired. Consequently, there are two different 
excitation energies, $\ep _{k,1}$ and $\ep _{k,2}$. 
At the Fermi surface, it costs no energy to excite them. 
More gapless modes due to the color and electric charge 
neutrality can be found in Ref.\ \cite{gapless}.

Here we study six different phases: 2SC phase, color-flavor locking (CFL) 
phase \cite{cfl}, color-spin locking and polar phases \cite{schaefer,when}. 
The first two are spin-0 phases while the last two are 
spin-1 phases. We solve the gap equation at zero temperature and
obtain the value of the gap function at the Fermi surface, $\phi_0$, 
for all cases in units of the 2SC gap value given by Eq.\ (\ref{phi02SC}): 
\be \label{ratio}
\frac{\phi_0}{\phi_0^{\rm 2SC}} = \exp(-d) \, 
\left( \lambda_1^{a_1} \, \lambda_2^{a_2} \right)^{-1/2}\,\, .
\ee 
This ratio is given in the sixth column of Table I 
of Ref.\ \cite{when}. The constant $d$ appears in spin-1 phases 
and originates from subleading contributions and leads to a tremendous
suppression of the gap by factors $e^{-4.5} \simeq 10^{-2}$
to $e^{-6} \simeq 2.5\times 10^{-3}$ relative to the 
spin-zero gap \cite{ren,schaefer}. 

We also obtain the transition temperature $T_c$, where the 
color-superconducting condensate melts: 
\be
\label{Tc}
\frac{T_c}{\phi_0}=\frac{e^\gamma}{\pi} \, 
\left( \lambda_1^{a_1} \, \lambda_2^{a_2} \right)^{1/2} 
\simeq 0.57 \, \left( \lambda_1^{a_1} \, \lambda_2^{a_2} \right)^{1/2}\,\, ,
\ee
where $\gamma\simeq 0.577$ is the Euler-Mascheroni constant.
In the 2SC and polar phases, where there is only one
gapped quasiparticle excitation,
$( \lambda_1^{a_1} \, \lambda_2^{a_2})^{1/2} = 1$,
we recover the relation $T_c/\phi_0 \simeq 0.57$
well known from the BCS theory \cite{BCS}.
Its validity for QCD in the 2SC phase 
was first demonstrated in Refs.\ \cite{pis00, wang}.
In the CFL and CSL phases there are two distinct gapped quasiparticle
excitations, and consequently two gaps, 
$\sqrt{\lambda_1} \, \phi_0 = 2 \, \phi_0$
and $\sqrt{\lambda_2} \,\phi_0 = \phi_0$. 
The BCS relation $T_c/\phi_0 = e^\gamma/\pi$ is violated by the 
additional factor $( \lambda_1^{a_1} \, \lambda_2^{a_2} )^{1/2} > 1$.

We have calculated the polarization 
tensor $\Pi_{ab}^{\mu\nu}(P)$, $a,b=1,\ldots,8,\gamma$, 
for gluons and photons in different color-superconducting phases. 
We have explicitly computed its zero-energy, low-momentum limit 
for the 2SC, CFL, polar and CSL phases, 
which yield the Debye and Meissner masses \cite{schmitt1,schmitt2}. 
These masses determine the screening lengths of electric and 
magnetic fields. Parts of the results were already known in the literature, 
namely the gluon Debye and Meissner masses for the spin-0 2SC and CFL 
phases \cite{meissner2,son,litim}. Our result for the photon Debye mass
in the 2SC phase shows that the photon-gluon mass matrix is already diagonal 
and thus electric gluons do not mix with the photon.  
The masses for the spin-1 phases have been computed in Refs. 
\cite{schmitt1,schmitt2} for the first time. For the polar phase, 
we have shown that there is mixing between the 
magnetic gauge bosons but, as in the 2SC phase, no mixing of the electric 
gauge bosons. In a system of one quark flavor, this mixing leads to a
vanishing Meissner mass. However, for more than one quark flavor, we have 
shown that, if the electric charges of the quark flavors are not identical, 
there is an electromagnetic Meissner effect in the polar phase, contrary 
to both considered spin-0 phases. For the CSL phase, we have
found a remarkable result that, for any number of flavors, neither
electric nor magnetic gauge fields are mixed. Since there is no 
vanishing eigenvalue of $\Pi^{\mu\nu}_{ab}(0)$, all eight gluons and the 
photon (electric as well as magnetic modes) become massive and 
there is an electromagnetic Meissner effect. For the polar phase, although 
there is a masssless new photon for each flavor, 
the electromagnetic Meissner effect still exists for a 
color superconductor with three flavors of quarks 
because there is no unique mixing angle for the new photon.  
We argued that, in spite of a suppression of the gap by 
three orders of magnitude compared to the spin-0 gaps \cite{schaefer,when},
spin-1 gaps might be preferred in a charge-neutral system. The reason is 
that a mismatch of the Fermi surfaces of different quark flavors
has no effect on the spin-1 phases, where quarks of the same flavor
form Cooper pairs. Therefore we found that a compact stellar object 
with a core consisting of quark matter in a 
spin-one color-superconducting state 
is, with respect to its electromagnetic properties, 
different from an ordinary neutron star: 
a spin-one color superconductor is an electromagnetic
superconductor of type I, while an ordinary neutron star is 
commonly believed to be of type II.
We note that a type-I superconductor could provide one
possible explanation for
the observation of pulsars with precession periods
of order 1 year \cite{link}.

We also derived a generalized Ward identity from QCD for 
dense, color-superconducting quark matter \cite{hwr,kobes,gerhold}.
The identity implies that, on the quasi-particle mass shell, 
the gap function and the 
quasi-particle dispersion relation are independent of the gauge 
parameter in covariant gauge up to subleading order.
We have shown that, to subleading order,
the gauge dependence of the quark self-energy arising
from the {\em gauge-dependent part\/} of the {\em gluon propagator\/} 
vanishes on the mass shell.
In principle, however, other gauge-dependent terms arise
from the gauge dependence of the {\em full vertex\/} and
of the {\em full quark propagator\/}, when
combined with the {\em physical part\/} of the gluon propagator.
We will show that these two cases do not provide additional 
subleading contribution to the gap function on the 
mass-shell in our future work. 
Our result shows that in order to obtain a gauge-independent gap
function up to subleading order, one has to use the full vertex 
as well as the full fermion propagator in the Nambu-Gor'kov basis. 
A consequence is that the prefactor $\exp(3 \xi/2)$ to the gap
function found in the mean-field approximation \cite{miransky,hong03}
will be removed by contributions from the full $qqg$ vertex
when taking the gap function on the quasi-particle mass shell.
An explicit diagrammatic proof of this statement will
be presented elsewhere.

\ack
The author thanks R. Pisarski and K. Rajagopal 
for helpful discussions during the conference. The work is 
supported by Alexander von Humboldt-Foundation, GSI Darmstadt and BMBF.

\end{document}